\documentclass[12pt]{iopart}
\usepackage{graphicx,cite}
\usepackage{amssymb}

\begin{document}

\title[Adiabatic compression of BEC solitons]
{Adiabatic Compression of Soliton Matter Waves}

\author{F. Kh. Abdullaev\dag \, and Mario Salerno \ddag}

\address{\dag Physical-Technical Institute, Uzbek Academy
of Sciences, 2-b, Mavlyanov str., 700084 Tashkent,Uzbekistan.}

\address{\ddag  Dipartimento di Fisica "E.R. Caianiello",
Universit\'a di Salerno, I-84081 Baronissi (SA), Italy, and
Istituto Nazionale di Fisica della Materia (INFM), Unit\'a di
Salerno, Italy.}

\date{\today}

\eads{\mailto{fatkh@physic.uzsci.net},
      \mailto{salerno@sa.infn.it}}

\begin{abstract}
The  evolution of atomic solitary waves in Bose-Einstein
condensate (BEC) under adiabatic changes of the atomic scattering
length is investigated. The variations of amplitude, width, and
velocity of soliton are found for both spatial and time adiabatic
variations. The possibility to use these variations to  compress
solitons up to very high local matter densities is shown both in
absence and in presence of a parabolic confining potential.
\end{abstract}

\pacs{
03.75.Fi, 
03.75.-b, 
05.45.Yv  
}

\submitto{\JPB}

\section{Introduction}
\label{introduction}

The discovery of Bose-Einstein condenstates (BEC) in atomic vapors
of alkali metals has opened new possibilities to observe matter
waves of soliton type. These excitations were first discovered in
BEC with repulsive interatomic interactions,  in the form of dark
solitons \cite{Den}, and more recently, in BEC with attractive
interatomic interactions, in the form of bright solitons moving on
zero backgrounds \cite{Strecker,Khaykovich}. Since soliton matter
waves are of primary importance for developing concrete
applications, it is of interest to devise methods which allow to
control them. One possibility is to vary the atomic scattering
length by means of external magnetic fields, i.e. by using
Feshbach resonances. In Ref.\cite{Castin}, it was shown that an
abrupt change in time of the atomic scattering length, can lead to
the splitting of the soliton  with the generation of new solitons.
The fragmentation of the wave obviously decreases the number of
atoms contained in the original pulse, this being undesirable for
applications such as atom lasers.

\noindent In the present paper we suggest to use {\it adiabatic}
variations of the atomic scattering length, both in time and in
space, as an effective way for controlling soliton's parameters
and to induce changes in their shape which could be useful for
applications. In contrast to abrupt variations, adiabatic changes
make possible to preserve the integrity of the soliton (no
splitting occurs), this leading for bright solitons to the
compression of the pulse with the increase of the matter density,
and for dark solitons to the compression of the hole with a
decrease of the background level (matter from background move
inside the hole). These phenomena are shown to exist both in
presence and in absence of a parabolic confining potential. In the
last case we derive analytical equations for soliton parameters in
terms of adiabatic invariants and soliton perturbation theory. The
results of this analysis is found in good agreement with direct
integration of the 1D Gross-Pitaevskii equation (GPE). The effect
of a confining parabolic potential on the  adiabatic dynamics of
the soliton is also investigated by numerical simulations. We find
that, for bright solitons, except for the oscillatory motion
around the bottom of the trap, the phenomena of pulse compression
is practically the same as in absence of the trap (this is
particularly true for solitons initially at rest in the bottom of
the trap). For dark soliton the phenomenon is only qualitatively
preserved, due to the boundaries effects introduced by the trap.
The possibility to compress BEC solitons could be an experimental
tool to investigate the range of validity of the 1D GPE. Since the
quasi one dimensional regime is valid for low densities, it would
be indeed interesting to see how far one can compress a soliton in
a real experiment by means of adiabatic changes of the scattering
length. Effects of adiabatic perturbations on soliton dynamics
were also investigated in the context of Josephson junctions
\cite{Salerno} and in nonlinear optics \cite{Anderson}.  In
contrast with Josephson junctions and optical fibers, which
require structural changes or preparation of new samples, the
study of adiabatic nonlinear perturbations on BEC solitons appears
more natural and easy to perform, since the strength of the
nonlinear interaction can be changed by using only external
fields. We also remark that soliton dynamics in a quasi
one-dimensional BEC under time-dependent {\it linear} potential
was recently  studied in \cite{Nis}.

The paper is organized as follows. In Section 1 we perform an
analytical study of the effects on bright and dark BEC solitons
induced by spatial and temporal  adiabatic variations of the
scattering length. For temporal variations we use both a
variational approach and perturbation theory to obtain equations
for soliton's parameters both in presence and absence of a
confining parabolic potential. In the former case the problem can
be linked to an adiabatic invariant of the Kepler problem with
results depending only on  adiabaticity (smoothness) but not on
the smallness of the perturbation. For spatial variations we use
standard soliton's perturbation theory. In Section 2 we compare
the results of this analysis with direct numerical integrations of
the full 1D GPE both in absence and in presence of an harmonic
trap. In the last section  we summarize the main results of the
paper.

\section{Analysis}

\noindent The dynamics of a dilute trapped BEC is described by the
Gross-Pitaevskii equation (GPE)
\begin{equation}\label{GP0}
i\hbar\phi_{t}(r,t) = -\frac{\hbar^{2}}{2 m}\Delta\phi(r,t) +
V_{tr}(r,t)\phi(r,t) + \Gamma |\phi|^{2}\phi(r,t),
\end{equation}
with $\Gamma=\frac{4\pi\hbar^{2}a_s}{m}$, $m$ being the atomic
mass, $a_s$ the scattering length, and   $V_{tr}
=\frac{1}{2}m(\omega_{1}^{2}x^2 + \omega_{2}^{2}y^2 +
\omega_{3}z^{2})$ represents the harmonic trap. The quasi 1D
geometry corresponds to the case $\omega_{1}^{2} <<
\omega_{2,3}^{2}$. To model adiabatic variation of the atomic
scattering in 1D (cigar shaped) BEC we consider the  following
normalized Gross-Pitaevskii equation \cite{Garcia,Rein}
\begin{equation}\label{GP1}
i \psi_t + \psi_{xx} + \sigma\gamma(x,t) \psi|\psi|^{2}-\omega^2
x^2 \psi =0,
\end{equation}
where $\psi$ is the ground state wavefunction of the condensate,
$\gamma(x,t)$ is a slowly varying function of space and time,
$\sigma \pm 1$ corresponds to the case of negative and positive
scattering length $a_s$ ($\omega$ denotes the longitudinal
frequency of the trap). In Eq. (\ref{GP1}) the space has been
normalized with respect to the healing length $\xi = 1/\sqrt{8\pi
\rho |a_{s}|}$ ($\rho$ is the atomic density), while the time with
respect to $t_{0} = m\xi^2/\hbar$. We remark that the 1D
approximation is valid for the number of atoms $N <
l_{r}/|a_{s}|$, where $l_{r} =\sqrt{\hbar/(m\omega_{2,3})}$ is the
transverse oscillator length. It is also worth to mention that in
the strong interaction limit $a_{s}N|\psi|^{2}>> 1$,  ($a_s>0$),
the effective nonlinearity is of the type $|\psi|\psi$ \cite{Sal}
so that deviations from Eq. (\ref{GP1}) occur (this will not be
considered in this paper). Although the analysis can be performed
for a generic smooth functions $\gamma(x,t)$, we shall restrict to
the limiting cases: $\gamma \equiv \gamma(t)$, and $\gamma \equiv
\gamma(x)$, the former being experimentally more easy to realize.

\subsection{{\it Case $\gamma \equiv \gamma(t)$}}
To study adiabatic temporal variations of the scattering length,
{\it case $\gamma \equiv \gamma(t)$}, we remark that, due to the
number of atoms conservation, it is natural to look for solution
of Eq. (\ref{GP1}) with $\sigma=+1$ (negative scattering length)
in the form of a bright soliton with time dependent parameters
\begin{equation}
\psi(x,t) =
A(t)\mbox{sech}(\frac{x-\zeta(t)}{a(t)})e^{ib(t)(x-\zeta(t))^{2} +
iC(x-\zeta(t))+ i\phi(t)},
\end{equation}
where $A(t)$, $a(t)$, $\phi(t)$, $\zeta(t)$, $C(t)$, $b(t)$,
denote, respectively, amplitude, width, phase, center, velocity
and "chirp" oscillation of the perturbed soliton. Taking these
parameters (with their derivatives), as collective coordinates for
the soliton, one can derive their time evolution from the
Euler-Lagrange equations with respect to the space averaged
Lagrangian $\bar{L} =\int L(x,t) dx$ of Eq. (\ref{GP1})
\cite{Malomed1}. This gives for the center of mass: $\zeta_{tt} +
4\omega^{2}\zeta =0$, and for the width \cite{Turitsyn}:
\begin{equation}
\label{eqwidth} a_{tt} + 4\omega^{2}a = \frac{16}{\pi^2 a^3}
-\frac{4\gamma(t)N}{\pi^{2}a^{2}}.
\end{equation}
Notice that Eq. (\ref{eqwidth}) has the form of a perturbed Kepler
problem \cite{Abdullaev}. This analogy can be used to reformulate
the problem in terms of the evolution of a unit mass moving in an
adiabatically varying Kepler potential. Since action-angle
variables for the unperturbed case are known, one can describe the
adiabatic change of the soliton width by means of the adiabatic
invariant of the Kepler problem. As the result we find for
$\omega^{2} < 1$ that
\begin{equation}a \approx
\frac{4}{\gamma(t)N}[1 - \frac{64\omega^{2}}{(\gamma(t)N)^{4}}].
\label{approx}
\end{equation}
We see that for $\omega=0$ one recovers the law $a =
4/(\gamma(t)N)$ known from nonlinear fiber optics. In this case,
one can also show that the frequency of the width oscillation is
given by $\omega_{a} \approx \gamma^{2}(t)N^2/4\pi$. The analysis
can be easily extended to two dimensions (2D) for the radially
symmetric GPE. In this case the variational equation for the width
becomes $a_{tt} + 2\omega^{2} a = \frac{2(2 - N_{1}\gamma_{1}(t))}{a^{3}}$, where
$N_{1} =N/(2\pi), \gamma_{1} =\gamma/(\sqrt{2}\pi l_{x}) $. Applying
the above arguments, one finds that the width of a 2D soliton
changes as $a \approx (2 - N_{1}\gamma_{1}(t))^{1/4}/\omega^{1/2}$. Eq. (\ref{approx})
predicts for small $\omega$ a dependence $1/\gamma$ of the soliton
width . This result can also be obtained by perturbation theory.
To this end we assume in the following that the range of variation
of the potential is very large in comparison with the size of the
condensate, so that we can neglect the $\omega$ term in Eq.
(\ref{GP1}) (this is particularly appropriate for static bright
solitons located at the bottom of the trap, as we will show
later). In this case the transformation $\psi = u/\sqrt{\gamma}$
(for $\gamma>0$), allows to rewrite Eq. (\ref{GP1}) in the form of
a perturbed nonlinear Schr\"odinger equation
\begin{equation}\label{GP2}
iu_{t} + u_{xx} + |u|^{2}u = i(\ln\sqrt{\gamma})_{t}u +
(\ln(\gamma))_{x}u_{x} + 0(\frac{T_{s}^2}{T^2},
\frac{L_{s}^2}{L^2}) = R(u),
\end{equation}
where $T$ and $L$ are, respectively, the temporal and spatial
characteristic scales of the inhomogeneity, while $T_{s},L_{s}$
are characteristic time and space soliton scales, assumed to be
much smaller than the corresponding  inhomogeneity scales. In the
case $\gamma_{x} =0$, the right hand side of Eq. (\ref{GP2})
represents a small linear time dependent amplification (damping).
If we consider the initial condition as a single soliton
\begin{equation}\label{sol}
u(x,t) = \sqrt{2}A(t)\mbox{sech}(A(t)(x-\zeta))e^{i\Theta(t)},
\;\;\; \Theta = k(t) + C(t)(x -\zeta(t)),
\end{equation}
then, using the equation for the energy $N = \int |u|^2 dx$, we
find $A = A_{0}\gamma(t)$ this giving, for the amplitude and width
of the soliton, $A_{\psi} = A_{0}\sqrt{\gamma}, a =
1/(A_{0}\gamma)$, in agreement with the variational approach.
\begin{figure}[htb]
\centerline{
\includegraphics[width=5cm,height=7.5cm,angle=270,clip]{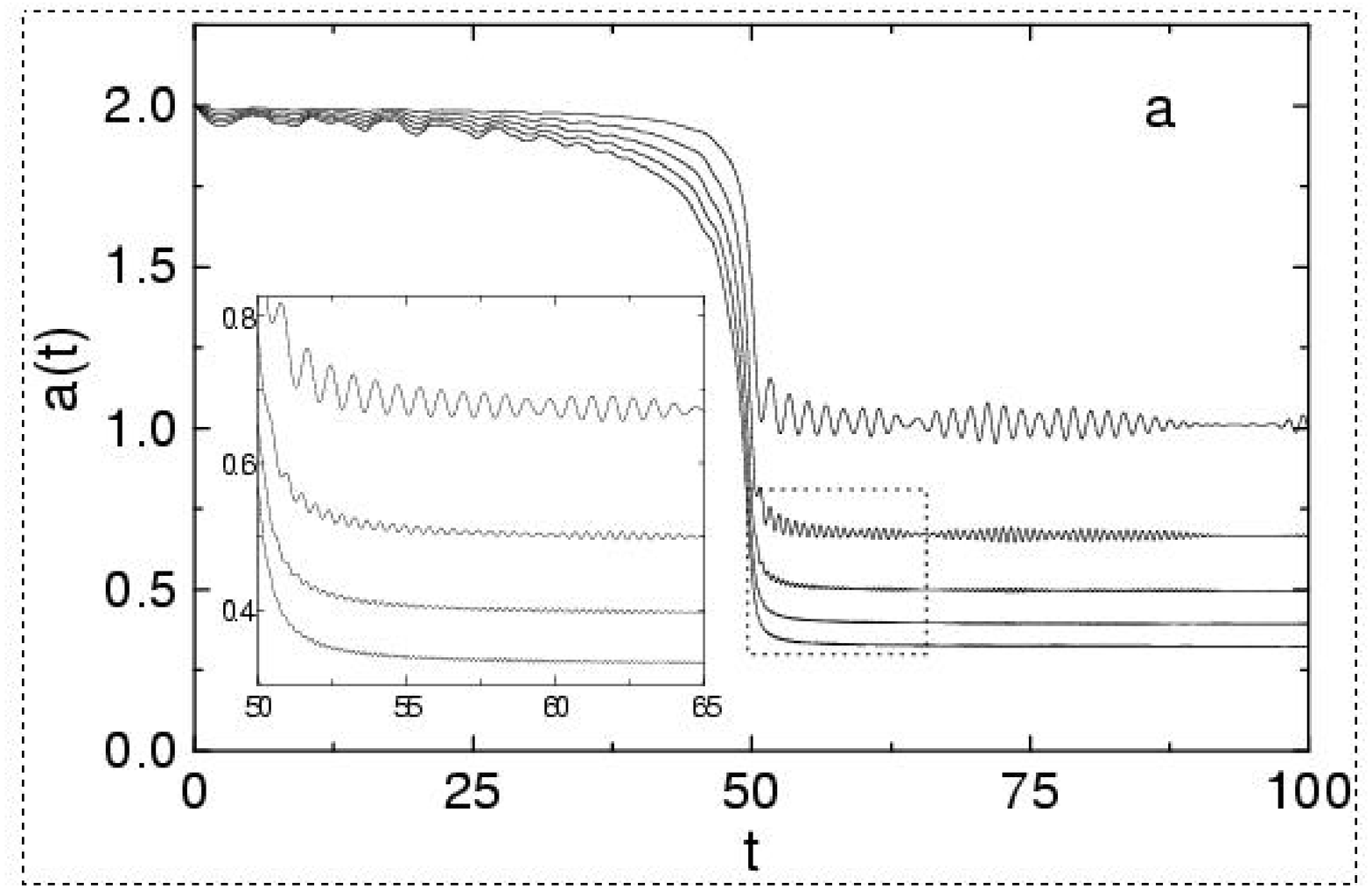} \quad
\includegraphics[width=5cm,height=7.5cm,angle=270,clip]{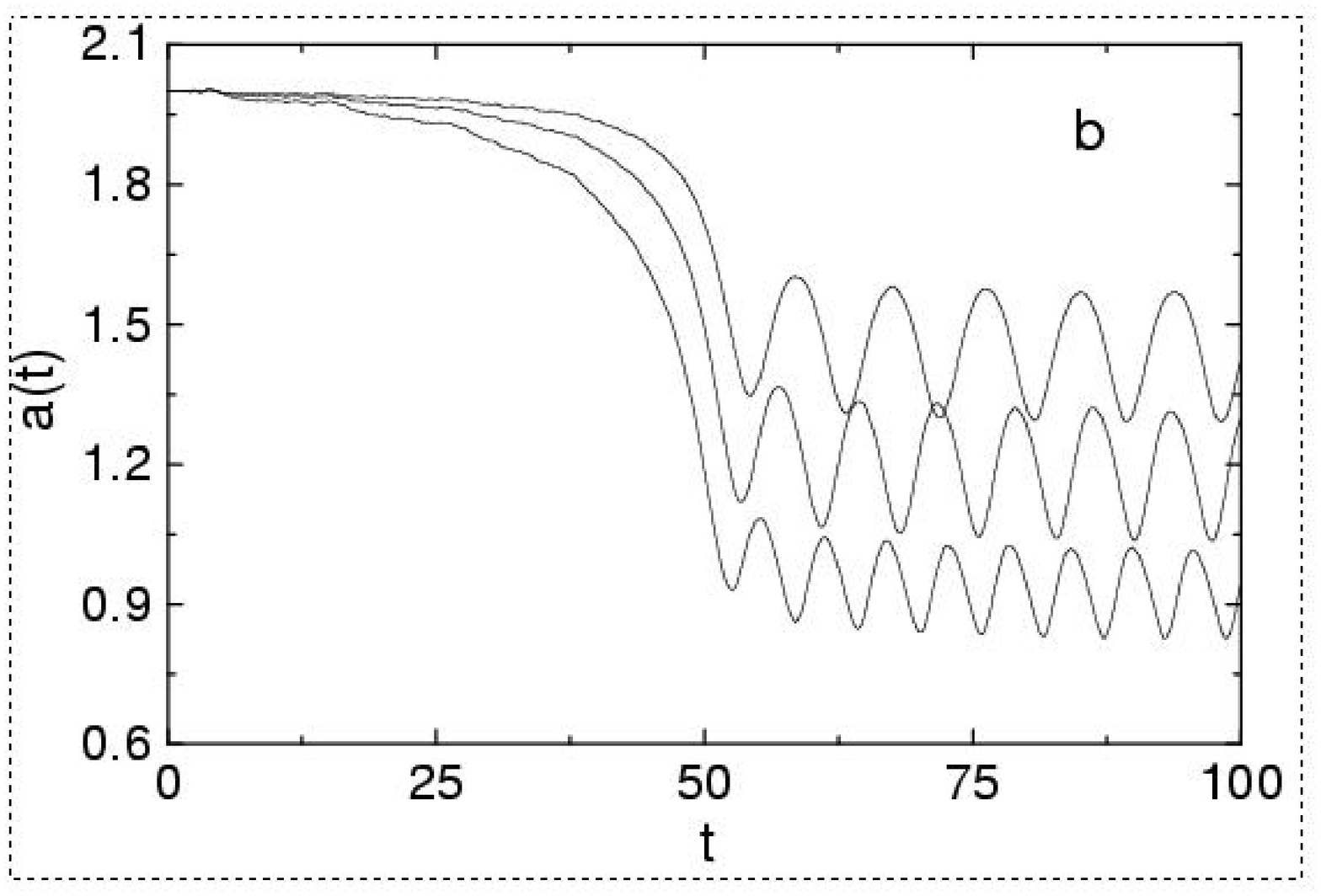}}
\caption{a) Soliton's width vs. time for different values of
$A_{s}$. Panel a) refers to the bright soliton case ($\sigma=1$)
The values of $A_s$ are stepped by 2, from $A_s=2$ (top curve) to
$A_s=10$ (bottom curve). The other parameters are $s=0.5,
T_f=100$, $\sigma=1$, $v_s=0$. An enlargement of the broken line
box, showing rapidly decreasing oscillations for increasing $A_s$
values, is reported in the inset. Panel b) refers to the dark
soliton case  ($\sigma=-1$ for $A_s$ values $2, 4 , 8,$ (from top
to bottom). The other parameters are $s=0.1, T_f=100$, $v_s=0$.}
\label{fig1}
\end{figure}
\noindent From this we conclude that an adiabatic increase of the
scattering length can be used to narrowing the width of a bright
soliton matter wave. Soliton compression phenomena induced by
linear damping amplification are also known from nonlinear optics
\cite{Anderson,Abdullaev}. A similar analysis can be performed for
the adiabatic evolution of a dark soliton ($\sigma=-1$ in
Eq.(\ref{GP1})). In this case, the adiabatic evolution of  the
dark soliton $\psi_{d}$ and of the background $u_{B}$ can be found
as
\begin{equation}
\psi_{d} =
u_{0}\tanh(\frac{u_{0}\sqrt{\gamma}}{\sqrt{2}}(x))e^{-i\theta}, \;
u_{B} = u_{0}\sqrt{\gamma}\exp(-i\theta), \;\ \theta =
-iu_{0}^{2}\int_{0}^{t}\gamma(t') dt',
\end{equation}
where $u_{0} =\sqrt{\mu}$, with $\mu$ the chemical potential of
the condensate (notice that this last expression for $u_{B}$ is
exact). From this expression we see that the width of the dark
soliton is changing according to $a_{d} =
a_{d0}/\sqrt{\gamma(t)}$, i.e. a factor $1/\sqrt{\gamma(t)}$ less
in comparison with bright solitons.

\subsection{{\it Case $\gamma \equiv \gamma(x)$}}

\noindent The perturbative approach can be used to study also the
case of  spatial variations of the scattering length  {\it i.e.
$\gamma \equiv \gamma(x)$}. We have from Eq. (\ref{GP2}): $R(u) =
F(x)u_{x}$, $F = (\ln \gamma(x))_{x}$, with $R(u)$ considered to
be a small perturbation.  Using the perturbation theory for
solitons \cite{Karpman}, we find that the equations for the
soliton's parameters  (\ref{sol}) are:
\begin{eqnarray}
A_{t} &=& AC\int_{-\infty}^{\infty}F(\frac{y}{A}+
\zeta)\mbox{sech}^{2}(y)
dy \approx 2 A C F(\zeta),\nonumber\\
 C_{t} &=& A^{2}\int_{-\infty}^{\infty}F(\frac{y}{A}+\zeta)(\mbox{sech}^{2}(y)-
\mbox{sech}^{4}(y))dy \approx \frac{2}{3}A^{2}F(\zeta),\\
 \zeta_{t} &=& 2C + \frac{C}{A}\int_{-\infty}^{\infty}F(\frac{y}{A}+\zeta)
y\mbox{sech}^{2}(y)dy \approx 2C(1 + O(\frac{1}{L^{2}})) \nonumber
.\label{csit}
\end{eqnarray}
From these equations it follows that ($F(-\infty) = 1$)
\begin{equation}
A = A_{0}\gamma(\zeta), C_{fin} = \sqrt{C_{in}^2 +
\frac{1}{3}A_{0}^{2} (\gamma^{2}(\zeta)-\gamma_{ini}^2)}.
\label{cfin}
\end{equation}
\begin{figure}[htb]
\centerline{
\includegraphics[width=5cm,height=7.5cm,angle=270,clip]{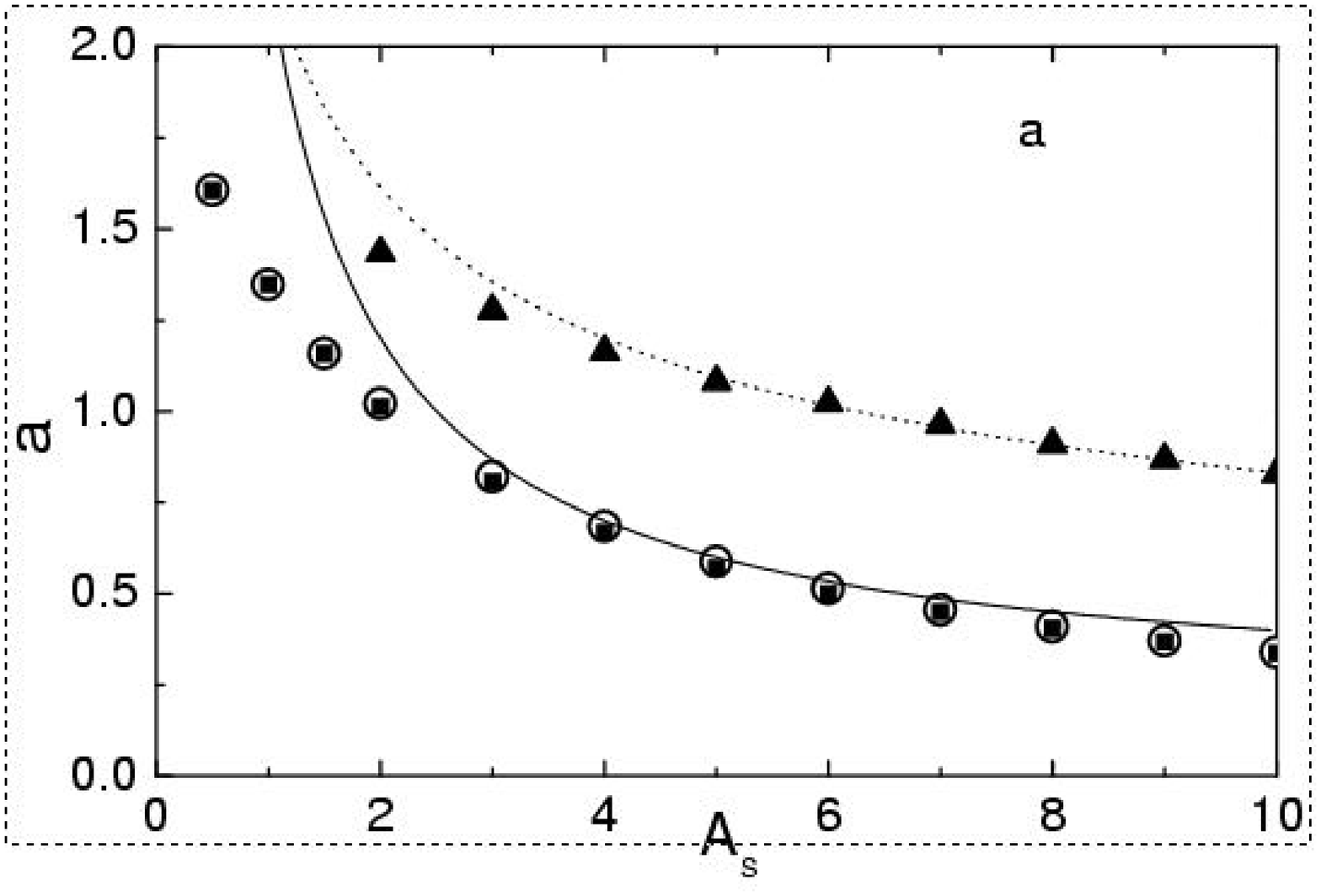} \quad
\includegraphics[width=5cm,height=7.5cm,angle=270,clip]{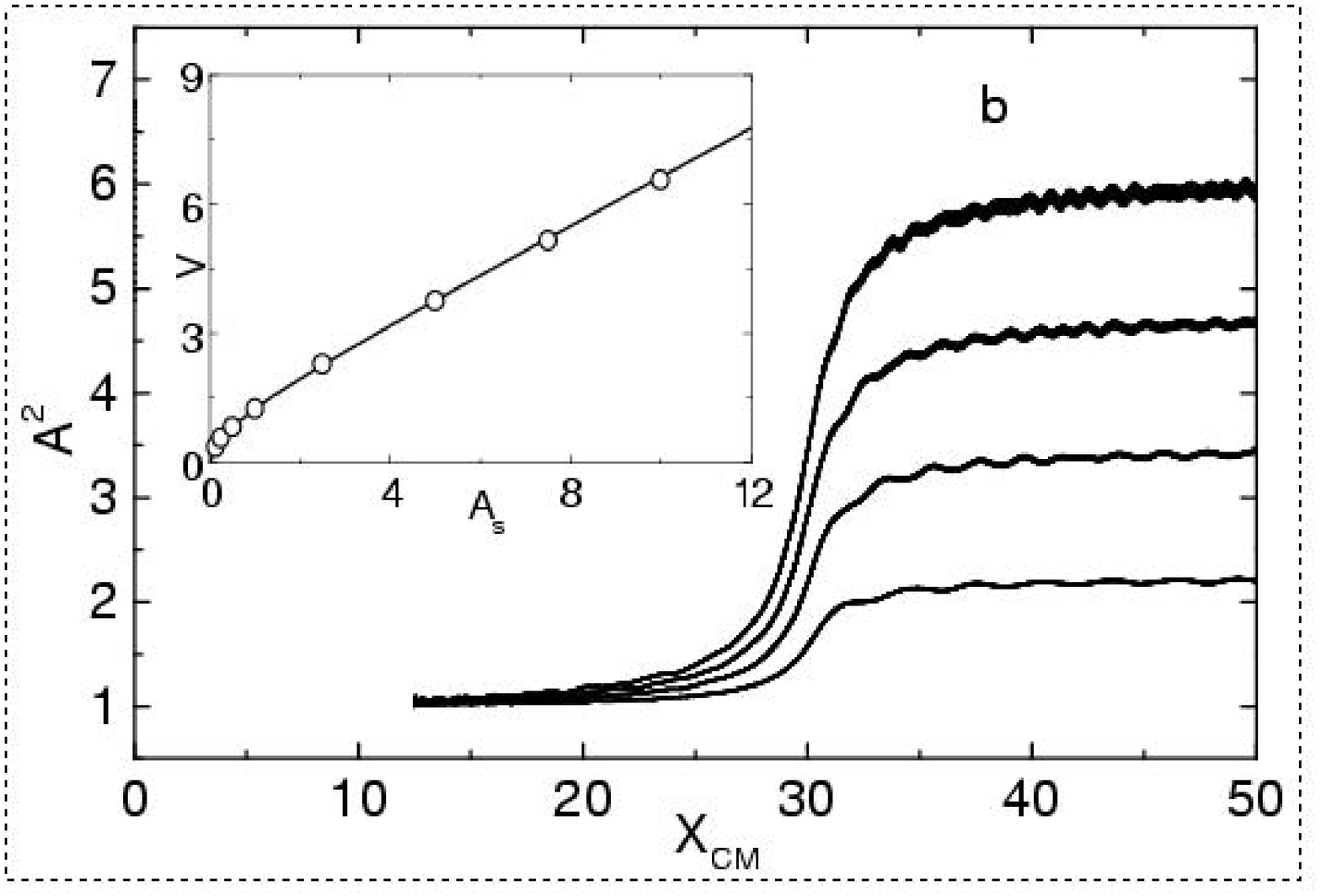}}
\caption{Panel a) Bright and dark soliton width versus step
amplitude $A_s$, for parameter values $s=0.1, T_f=100$, and
soliton velocity $v_s=0$. The triangles and squares denote
numerical values in absence of trap potential for $\sigma=1$
(bright case) and $\sigma=-1$ (dark case), respectively. The open
circles denote numerical values for the bright case in presence of
a trap potential of frequency $\omega^2=0.01$ (the soliton is
taken at rest in the bottom of the trap). The continuous and
dashed curves are the analytical predictions in absence of trap
potential for, respectively, bright and dark soliton case. Panel
b) The squared amplitude of a bright soliton vs center of mass
position for different values of the amplitude $A_s$ of a
kink-like spatial inhomogeneity centered at x=30, given by Eq.
(\ref{adiavariation}). The curves, from bottom to top, refer to
$A_s=2.5, 5.0, 7.5, 10.0$, respectively. The other parameters are
$s=0.2, T_f=60, \sigma=1, A_0=1$. The soliton is initially at
rest, placed at position $x_{ini}=12.5$. In the inset we show the
soliton final velocity as a function of $A_s$. The open dots are
numerical values while the continuous curve is obtained from Eqs.
(\ref{csit},\ref{cfin}) as $V_{fin}=\frac{A_0}{\sqrt 3}
\sqrt{\gamma_{fin}^2-\gamma_{in}^2}$.} \label{fig2}
\end{figure}
In the next section we shall compare these predictions with direct
numerical integrations of Eq.(\ref{GP1}).

\section{Numerical results}

\noindent To check these results we have numerically integrated
Eq. (\ref{GP1}) with the function $\gamma (t)$ given by
\begin{equation}
\gamma(t)= 2 + A_s \left[\frac{1}{2} + \frac{1}{\pi} \tan^{-1}(s
\pi)(t-\frac{T_f}{2})\right] \label{adiavariation}
\end{equation}
This function models, for small $s$, an adiabatic change of the
scattering length, while for large $s$ it reduces to the step
function of amplitude $A_s$ centered at $T_f/2$ (to have the
function starting from $2$, the quantity $\delta=\gamma(0)-2$ must
be subtracted). In the following we shall use this function to
model also spatial variations of the scattering length.  In Fig.
\ref{fig1} we depict the width of a bright (panel a) and a dark
soliton (panel b) as a function of time for different values of
$A_{s}$, in the case $\omega=0$. From this figure we see that the
oscillations of the soliton width after the transition, are more
pronounced for dark solitons than for bright ones. This is a
consequence of the fact that in the dark case the background is
also excited by the perturbation and contribute for a relevant
part to the dynamics. In Fig. \ref{fig2}a we show the average
final value of the width of bright and dark solitons as a function
of $A_s$. It is evident that, for the same absolute variation of
the scattering length, bright solitons are more compressed than
dark ones, the difference being just a factor proportional to
$1/\sqrt{\gamma}$, as expected from our analysis. The agreement
with the theoretical predictions is indeed rather good especially
for higher values of $A_{s}$. Notice that in this figure the
influence of a parabolic trap on the compression phenomenon is
also shown for the case of a bright soliton initially at rest in
the bottom of the trap (open circles). We see that the results are
almost identical to those in absence of trap (this is quite
expected, since the pulse is localized in the bottom of a low
intensity trap). The panel (b) of the same figure shows the
amplitude of a bright soliton as a function of the position of the
center of mass for the case of a space dependent variation of the
scattering length in absence of the parabolic trap. The soliton,
is initially at rest, is sucked into the higher scattering length
region, and reaches a constant velocity after passing the
inhomogeneity. Also in this case we find an excellent agreement
with our analysis (see the inset of the figure).
\begin{figure}[tbp]
\centerline{
\includegraphics[width=6.0cm,height=6.8cm,clip]{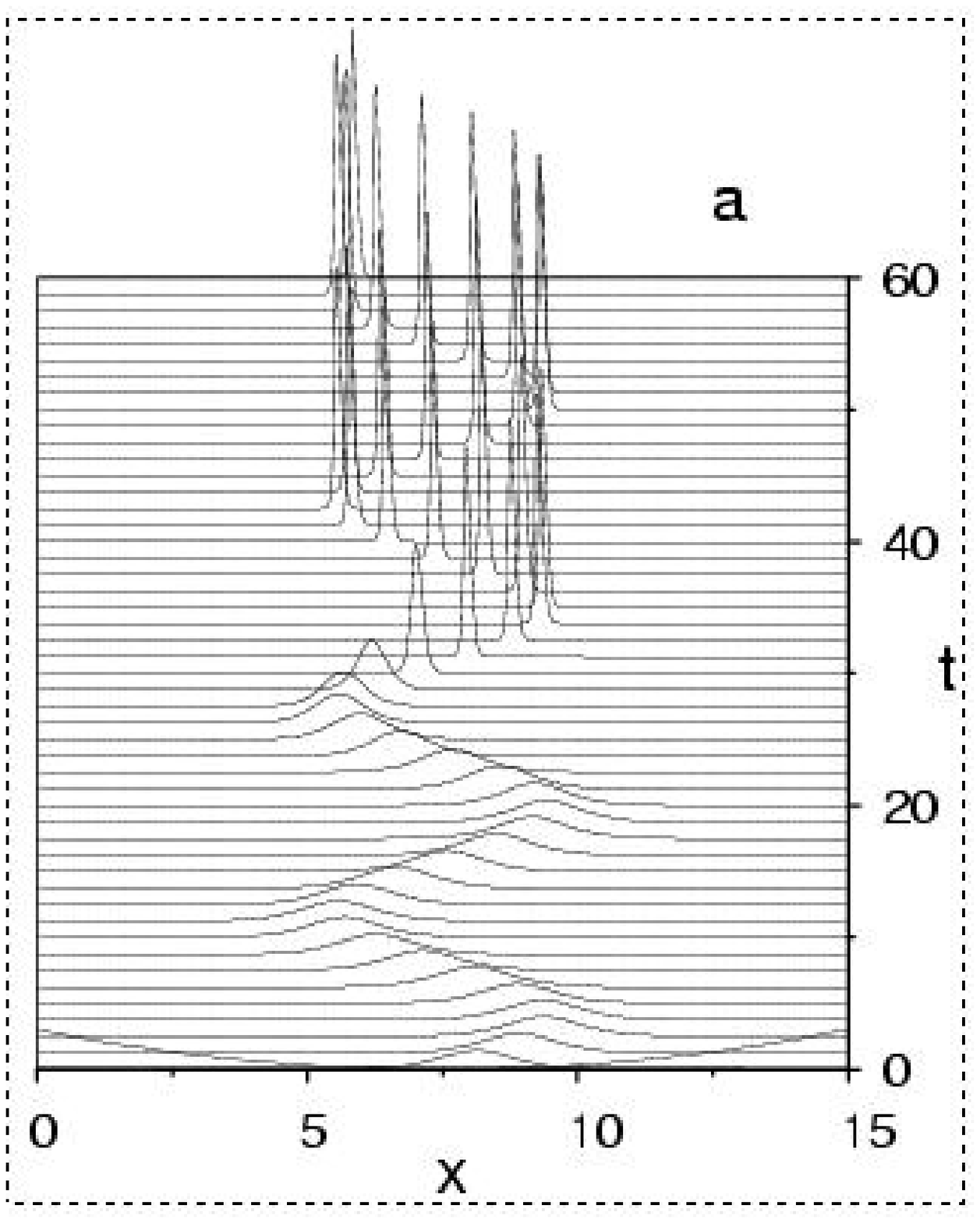} \quad
\includegraphics[width=6.0cm,height=6.cm,clip]{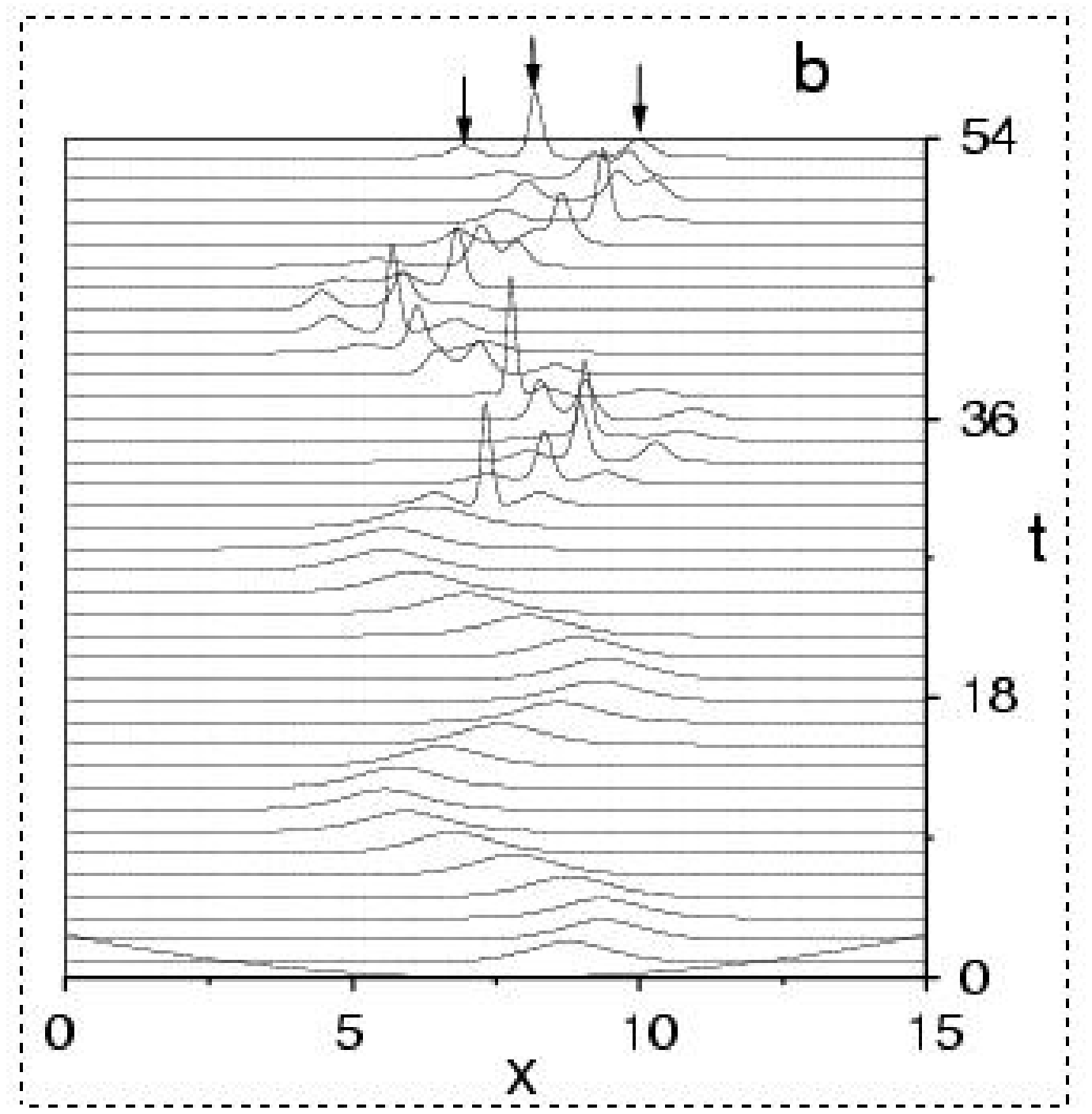}}
\centerline{
\includegraphics[width=6.0cm,height=6.cm,clip]{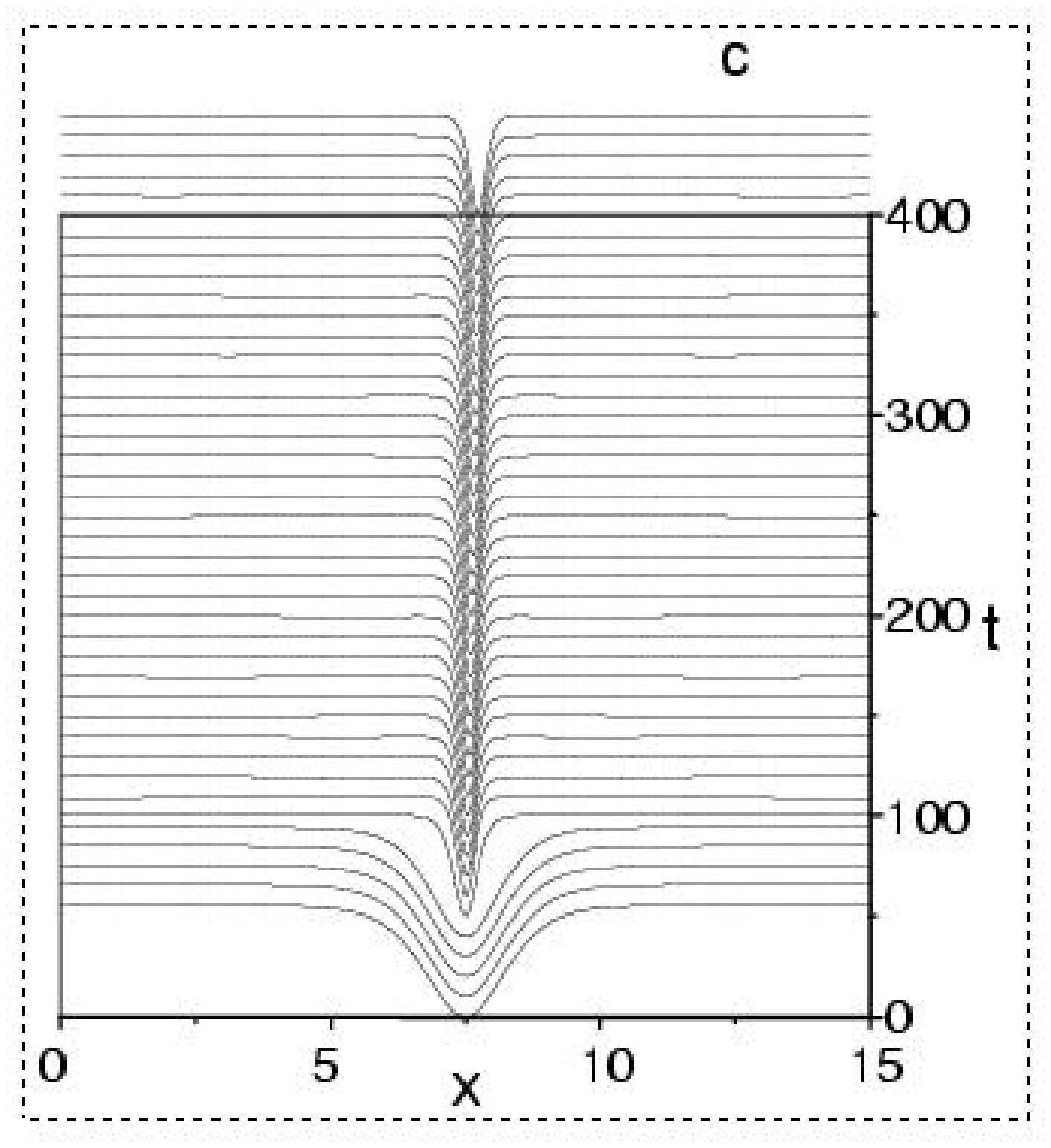} \quad
\includegraphics[width=6.0cm,height=6.cm,clip]{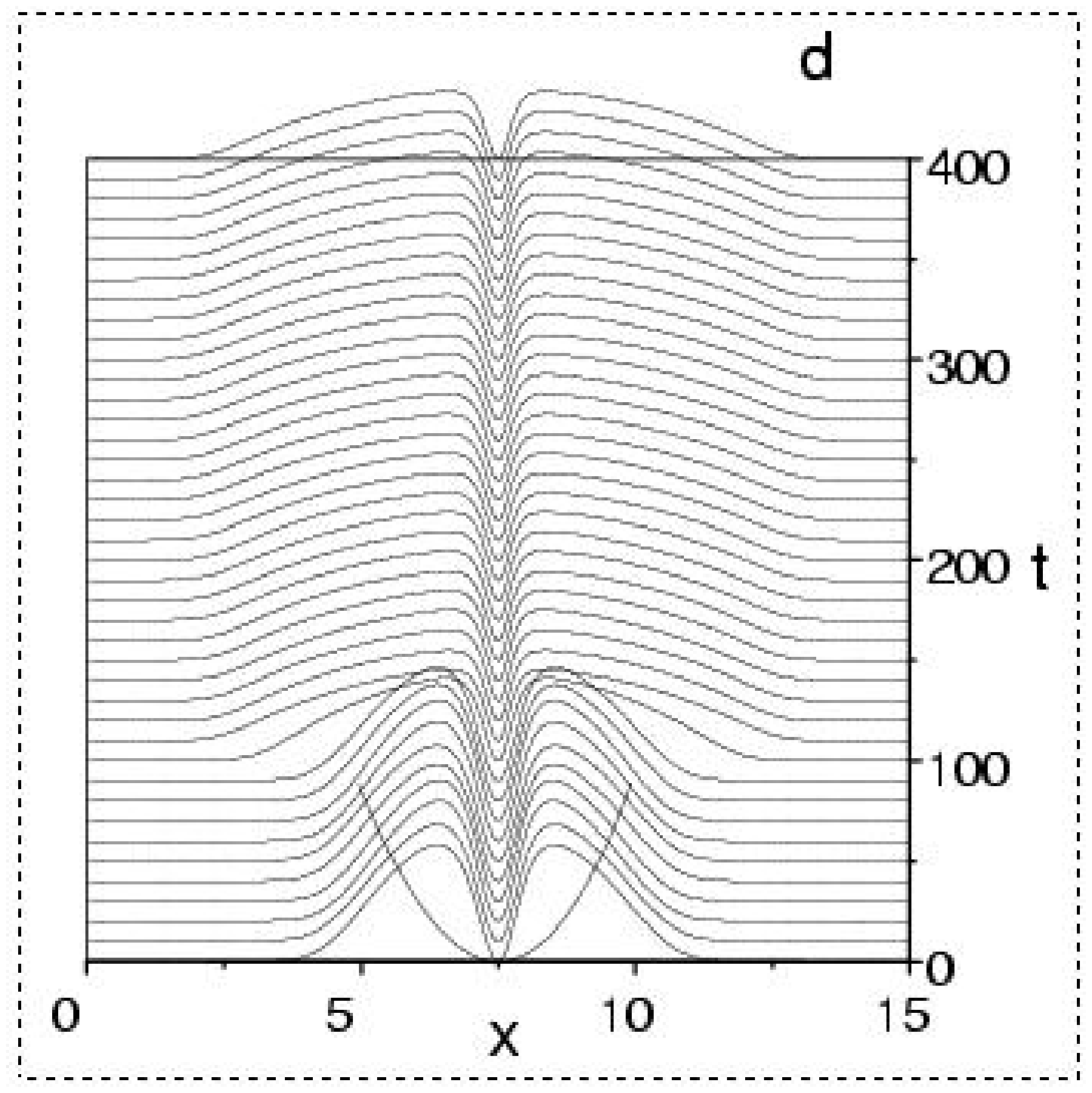}}
\caption{ Time evolution of bright and dark solitons in absence
and in presence of parabolic trap. Panels (a,b): Bright soliton
dynamics ($\sigma=1$) in presence of parabolic potential of
frequency $\omega^2=0.04$ subject to (a) an adiabatic change
$\gamma(t)$ of the scattering length with $A_s=25$, $s=0.5$, and
$T_f=50$; (b) a step like variation $\gamma(t)$ with $s=10^4$.
Other parameters are fixed as in panel (a). Panels (c,d): Dark
soliton dynamics $\sigma=-1$ in absence (c) and in presence (d) of
a parabolic trap of intensity $\omega^2=0.8$. In both cases the
initial dark profiles are at rest and subject to a $\gamma(t)$
variation with parameters $A_s=10$, $s=0.5$, and $T_f=200$. Notice
that the parabolic potential is also shown at $t=0$ to display the
scale in the z-direction.} \label{fig3}
\end{figure}

The effect of the parabolic trap on pulse compression is further
investigated in Fig. \ref{fig3}. In panels (a) and (b) of this
figure, we show the evolution of a bright soliton oscillating in
the trap and subject, respectively,  to adiabatic and abrupt
variation of the scattering length. We see that in contrast with
the adiabatic case, which always leads to compression, abrupt
variations of the scattering length may induce the splitting of
the soliton into sub-solitons (notice in panel (b), that three
solitons, indicated by the arrows,  are formed at time $t=54$).
The splitting phenomenon is similar to the one observed in absence
of parabolic potential \cite{Castin}. The number of sub-solitons
created can also be determined in analogy with this last case (we
omit details for brevity). Panels (c) and (d) of Fig. \ref{fig3},
show the time evolution of a dark soliton in absence and in
presence of a parabolic trap. We remark that, while in the bright
case the compression of the soliton is obviously accompanied by an
increase of its amplitude, in the dark case the amplitude cannot
decrease below zero, so that the compression must be  combined
with a decrease of the background level (the matter from the
background move inside the hole so to reduce its size). This is
particularly evident in panel c for the case $\omega=0$. We remark
that this case, although physically non realisable for BEC, it is
of interest for nonlinear optics, and shows the role played by the
confinement potential  in the case of BEC with positive scattering
lengths (it should be compared with panel(d) discussed below).

In analogy with bright solitons, one could expect that an abrupt
change in time of the scattering length may split a dark soliton
ground state into dark and grey sub-components if the amplitude of
the perturbation is big enough. This is indeed what we numerically
observed. For the case $\omega=0$, using the Inverse Scattering
Transform, one can predict that out of an initial profile of the
form $\psi(x,t=0) = b\tanh(c x)$, one black soliton and  a number
$N_{0}$ of pairs of grey solitons (with $N_{0} = n-1$, $n$ being
the integer part of the ratio $b/c$) \cite{Zhao,Tsoy}, should be
generated. By denoting with $\gamma_{1}$ and $\gamma_{2}$ the
initial and final values of $\gamma$, one can write  the initial
condition for the equation: $iv_{t} + v_{xx} -|v|^2 v =0$,  as
$v(0) = u_{0}\sqrt{\gamma_{2}}\tanh(u_{0}\sqrt{2\gamma_{1}}x)$,
with $v = \psi\sqrt{\gamma_{2}}$, and the  ratio $b/c$ equal to
$\sqrt{\gamma_{2}/2\gamma_{1}}$. We checked that this analysis
indeed gives correct predictions of splitting in absence of the
parabolic trap. The influence of the parabolic trap on a dark
soliton initially at the rest in its bottom, is reported in panel
(d) of Fig. \ref{fig3}. Although the phenomenon resemble the one
observed in absence of the trap (notice the decrease of the
background after the transition), the solution is more complicated
due to the boundary effects introduced by the trap. In particular
we see that the compression of the hole in the center is less
effective than for $\omega=0$, and the decrease of the background
is accompanied by an expansion of the condensate. This is a
consequence of the non constancy of the background (it obviously
must go to zero at large distances) which permits the matter to
escape up in the trap (the pressure on the hole is thus partially
lost in the expansion). The same phenomenon exists also for dark
solitons initially oscillating in the trap.  In analogy with the
case $\omega=0$, we also found that abrupt variations in presence
of parabolic potential induce dark soliton splitting.
\begin{figure}[htb]
\centerline{
\includegraphics[width=8cm,height=12.cm,angle=270,clip]{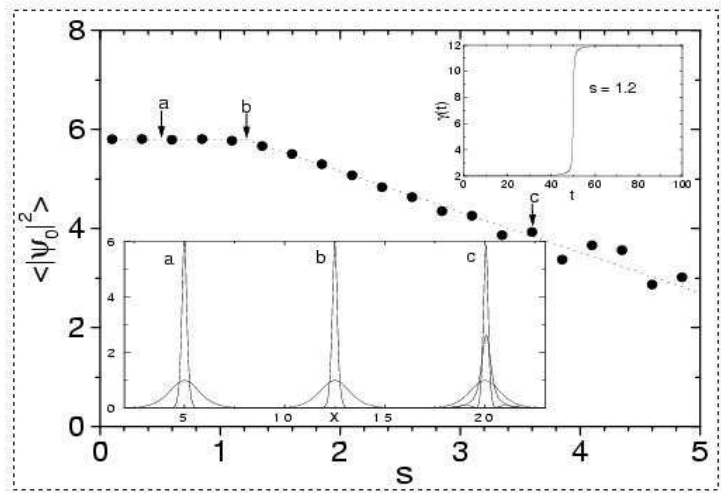}}
\caption{Time averaged amplitude of the soliton versus the
parameter $s$ of the scattering length function $\gamma(t)$. The
amplitude and the time of the variation are fixed as $A_{s}=10$
and $T_{f}=100$. Notice the crossover from adiabatic to abrupt
behavior occurring at $s\approx 1.2$ (dotted lines are drown to
guide the eyes). The inset at the top shows the shape of the
function $\gamma(t)$ at the crossover point (point b). The inset
at the bottom shows the initial and final profiles of the soliton
at the points $a, b, c,$ corresponding, respectively, to the
values, $s=0.5$, $s=1.2$, $s=3.6$. Soliton profiles corresponding
to points  a and c have been shifted by $\pm 7.5$ along the x axis
to avoid overlapping with the ones corresponding to point b. }
\label{fig4}
\end{figure}

In Fig. 4 we show the averaged amplitude of the soliton (i.e. the
time average of the modulo squared at the center of the profile)
versus the parameter $s$ controlling the rapidity of change of the
function $\gamma(t)$. From this figure we see that a crossover
from adiabatic to fast variations behavior occurs at $s\approx
1.2$. Notice that this crossover value of $s$ corresponds to a
variation function $\gamma(t)$ which is already quite rapid (see
top inset of the figure). The adiabatic region is characterized by
the fact that the final profiles of the compressed solitons are
almost independent on $s$ (they depend only on the amplitude of
the scattering length variation). This can be seen from the bottom
inset of Fig. \ref{fig4} in which the initial ($t=0$) and final
($t=100$) soliton profiles corresponding to points a and b in the
adiabatic region, are reported. Very little oscillations of the
soliton profile are found in these cases. On the contrary, the
non-adiabatic (or fast oscillation) region is characterized by the
presence of large oscillations of the soliton profile, induced by
the rapid variation of the scattering length. This is illustrated
in point c of the lower inset of Fig. \ref{fig4}, in which we
depicted, besides the initial profile at $t=0$, two snapshots of
the soliton (max and min of the profile) to evidenciate its
oscillations at the final time $T_f=100$. These oscillations
increase by increasing $s$ and eventually lead to the splitting of
the soliton as illustrated in panel b of  Fig. \ref{fig3} (the
splitting, however, depends also on the amplitude of the
scattering length variation, and usually occurs much above the
interface between adiabatic and fast oscillation behavior). From
this analysis we conclude that the adiabatic compression of matter
wave solitons is possible in a wide range of the parameter $s$
(i.e. $0<s \approx 1$) and is compatible with relatively fast
variations of the scattering length, as one can see from the top
inset of Fig. 4.

\section{Conclusions}

In conclusion, we have shown that adiabatic changes of the
scattering length, both in space and time, can be effectively used
to control parameters of BEC matter waves of soliton type. In
particular we showed the effect of pulse compression both on dark
and bright solitons in presence and absence of a confining
parabolic potential. The influence of the smoothness of the
transition was also considered. We found that deviations from
adiabatic predictions occur only for very large values of $s$,
i.e. for almost step-like variations. This implies that the
results of this paper are valid for generic smooth variations of
the scattering length and therefore are expected to be found in
real BEC matter waves experiments. The possibility of sharpening
bright and dark solitons by means of smooth variations of the
scattering length, beside providing another confirmation of the
soliton nature of the BEC ground state, may be very useful for
concrete applications. We hope that experiments in this direction
will be soon performed.

\ack We are grateful to V.V. Konotop for interesting discussions.
F.Kh.A. acknowledges the hospitality received at the Physics
Department of the University of Salerno, where this work was done,
and partial financial support from NATO-Linkage grant No.
PST.CLG.978177. M. S. thanks the European grant LOCNET no.
HPRN-CT-1999-00163 for partial support.


\section*{References}
\begin{thebibliography}{10}
\bibitem{Den}
Denschlag et al., {\it Science,} {\bf 287} (2000) 97.

\bibitem{Strecker}
Strecker K., Patridge G., Truscott A., and Hulet R., {\it Nature,} {\bf 417}
(2002) 150.

\bibitem{Khaykovich}
Khaykovich L., et al., {\it Science,} {\bf 296} (2002) 1290.


\bibitem{Castin}
Carr L.D. and  Castin Y., {\it Phys.Rev}, A, {\bf 66} (2002)
063602.

\bibitem{Salerno}
Salerno M., Samuelsen M.R.,  Lomdahl P.S.,  and  Olsen O.H., {\it
Phys.Lett.,} {\bf 108A } (1985) 241;\;\;\; Pagano S., Salerno M.,
and Samuelsen M.R., {\it Physica D,} {\bf 26}
 (1987) 396.

\bibitem{Anderson}
Quiroga-Teixero M.L. et al., {\it  JOSA B,} {\bf 13}  (1996) 687.

\bibitem{Nis}
Nistazakis H.E., et al., arXive:cond-mat/0211702 v1.

\bibitem{Garcia}
 Per\'{e}z-Garcia V.M., Michinel H., Herrero H., {\it Phys.Rev. A,} {\bf 57} (1998) 3837.

\bibitem{Rein}
Reinhardt W.P. and Clark C.W., {\it J.Phys.}, B, {\bf 30} (1997) L785.



\bibitem{Malomed1}
 Malomed B.A., In {\it "Progress in Optics",} Ed.E. Wolf, {\bf 43}, 71,
2002.

\bibitem{Turitsyn}
 Turitsyn S.K., {\it JETP Letters,} {\bf 65} (1997) 845.

\bibitem{Abdullaev}
 Abdullaev F.Kh. and  Caputo J.G., {\it Phys.Rev. E,} {\bf 58} (1998) 6637.

\bibitem{Sal}
Salasnich L.,  Parola A., and Reatto L., {\it Phys.Rev. A} (2002)
{\bf 043603}.

\bibitem{Karpman}
 Karpman V.I., {\it Phys.Scripta,} {\bf 20} (1979) 462.

\bibitem{Zhao}
 Zhao W. and Bourkoff E., {\it Opt.Lett,} {\bf 14} (1989) 703.

\bibitem{Tsoy}
Abdullaev F.Kh., Nurmanov N., and  Tsoy E.N., {\it Phys.Rev. E,} {\bf 56} (1997) 3638.

\end{thebibliography}
\end{document}